\documentstyle[epsf,prl,aps,floats]{revtex} 
 
\begin{document} 
\draft 
\wideabs{  
  
\title{``Linearized'' Dynamical Mean-Field Theory 
for the Mott-Hubbard transition} 
 
\author{R.\ Bulla$^{\rm (1)}$ and M.\ Potthoff$^{\rm (2)}$} 
\address{$^{\rm (1)}$Theoretische Physik III, 
Elektronische Korrelationen und Magnetismus, 
Universit\"at Augsburg, D-86135 Augsburg, Germany\\ 
$^{\rm (2)}$Theoretische Festk\"orperphysik,
Institut f\"ur Physik, Humboldt-Universit\"at zu Berlin, 
D-10115 Berlin, Germany} 
 
\maketitle 
 
\begin{abstract} 
The Mott-Hubbard metal-insulator transition is studied within a simplified 
version of the Dynamical Mean-Field Theory (DMFT) in which the coupling 
between the impurity level and the conduction band is approximated by a 
single pole at the Fermi energy. In this approach, the DMFT equations are 
linearized, and the value for the critical Coulomb repulsion $U_{\rm c}$ 
can be calculated analytically. For the symmetric single-band Hubbard model
at zero temperature,
the critical value is found to be given by 6 times the square root of the 
second moment of the free ($U=0$) density of states. This result 
is in good agreement with the numerical value obtained from the Projective 
Selfconsistent Method and recent Numerical Renormalization Group calculations
for the Bethe and the hypercubic lattice in infinite dimensions.
The generalization to more complicated lattices is discussed. 
The ``linearized DMFT'' yields plausible results for the 
complete geometry dependence of the critical interaction.
\end{abstract} 
 
 
} 
 
\section{introduction} 
 
The correlation-induced 
transition from a paramagnetic metal to a paramagnetic insulator 
(the Mott-Hubbard transition \cite{Mott,BUCH}) has been intensively studied 
within the single-band Hubbard model \cite{Hubbard,Gut,Kan}:
\begin{equation} 
   H = \sum_{\langle ij \rangle \sigma} 
         t_{ij} c^\dagger_{i\sigma} c_{j\sigma} + 
         U\sum_i c^\dagger_{i\uparrow} c_{i\uparrow} 
            c^\dagger_{i\downarrow} c_{i\downarrow}  \: . 
\label{eq:H} 
\end{equation} 
The model describes conduction electrons 
with spin $\sigma$ on a lattice with nearest-neighbor hopping 
matrix element $t_{ij}$ and a local Coulomb repulsion $U$. 
One of the first approaches to describe the metal-insulator 
transition in the half-filled Hubbard model has been the Hubbard-III 
approximation \cite{Hub64b}. The alloy-analogy solution predicts a 
splitting of the density of states in upper and lower Hubbard bands 
for large values of $U$. 
On decreasing $U$, the insulator-to-metal transition occurs 
when the Hubbard bands start to overlap. The 
critical interaction is approximately given by the free bandwidth: 
$U_{\rm c} \approx W$. 
The Hubbard-III approximation, however, fails to describe the Fermi-liquid 
properties in the metallic phase. 

Later, the Mott-Hubbard transition has been described within the 
Gutzwiller variational approach by Brinkman and Rice \cite{BR70}.
Starting from the metallic side, the transition 
is marked by a diverging effective mass. The critical interaction
is found to be $U_{\rm c}=-16 e_0$ where $e_0$ is the kinetic energy of
the half-filled band per particle for $U=0$. 
The Brinkman-Rice approach, however, fails to 
describe the insulating phase above $U_{\rm c}$.

A Dynamical Mean Field Theory (DMFT), which becomes exact in the limit
of infinite spatial dimensions, has been developed for the Hubbard model 
\cite{MVPRLdinfty,PJF,Georges}.
The DMFT is able 
to yield a consistent description of the metallic Fermi liquid
for weak coupling as well as of the Mott-Hubbard insulator for strong 
coupling.
In practice, however, the solution of the mean-field equations is by 
no means a trivial task.
In particular, for $U\mapsto U_{\rm c}$ problems may arise since the 
``mean field'' $\Delta(\omega)$ which has to be determined 
self-consistently, develops a strong frequency dependence on a 
vanishingly small energy scale.

The first calculations for temperature $T=0$ were performed using the 
so-called Iterated Perturbation Theory (IPT) \cite{Georges}. 
Within the IPT the highly correlated Fermi liquid for $U\mapsto U_{\rm c}$
is characterized by a narrow quasiparticle peak that is well isolated
from the Hubbard bands. As a consequence the insulating gap appears
to open discontinuously at the metal-insulator transition.
These characteristics of the transition have been questioned by various 
authors \cite{Kehrein,David,NG} so that the issue of the metal-insulator 
transition for $T\!=\!0$ (and also for $T>0$ \cite{Georges,Sch98}) cannot 
be regarded as settled at the moment.

Qualitatively, the IPT scenario for $T\!=\!0$ is corroborated by recent 
non-perturbative
calculations using the Numerical Renormalization Group (NRG) method 
\cite{BHP,NRG_new}. However, the critical value for the transition
is found to be significantly lower as compared to the IPT result. On the 
other hand, the NRG value for $U_{\rm c}$ is in remarkable agreement with 
the result of the Projective Self-consistent Method (PSCM) \cite{Moeller}.

The value of the critical interaction for the Mott-Hubbard 
transition is of great interest. Using the methods mentioned above, 
an approximate calculation of $U_{\rm c}$ is possible. This, however, 
represents a rather difficult numerical problem, the solution of which 
still depends on the approximation used. Within the framework of DMFT, 
an exact {\em analytical} result for the precise value of $U_{\rm c}$ 
is still missing. Even an approximate analytical expression is not 
available up to now. 

With the present paper we propose a simplified treatment of the 
mean-field equations (``linearized'' DMFT) which allows to obtain 
an explicit expression for $U_{\rm c}$ at zero temperature. 
A fully numerical treatment of the DMFT would leave us with a mere 
number for $U_{\rm c}$ and would hardly show up the characteristic 
trends for different geometries unless a large number of cases were 
studied. Contrary, the linearized DMFT is able to yield at once the 
complete geometry dependence of $U_{\rm c}$. In our opinion this outweights 
the necessity for further approximations. 

The main idea of the linearized DMFT is to approximate the hybridization 
function for the coupling between the impurity level and the conduction 
band by a single pole. This is detailed in Sec.\ II.
The reliability of the new approach is estimated by comparing the 
analytical results for the Bethe and the hypercubic lattice in infinite
dimensions with the available numerical values from the PSCM and the NRG 
in Sec.\ III A. A satisfactory agreement is found. In the following
we then demonstrate the predictive power of the approach. The geometry 
dependence of $U_{\rm c}$ is derived for a number of more complicated lattice 
structures: inhomogeneous Bethe lattices (III B and C) and hypercubic 
films in infinite and finite dimensions (III D). Finally, in Sec.\ III E 
we discuss a first correction beyond the linearized theory.
Sec.\ IV summarizes the main results.

\section{Linearized Dynamical Mean-Field Theory} 

A characteristic feature of the metal-insulator transition is that the 
quasiparticle peak appears to be isolated from the upper and the lower 
Hubbard band for $U \mapsto U_{\rm c}$ and $T=0$. 
Whether or not there is a real gap, i.~e.\ zero spectral weight
between the quasiparticle peak and the Hubbard bands, is difficult to
decide with any numerical method but not very important for the present
approach. Essentially, our approach is based on two approximations:

(i) We assume that in the limit $U \mapsto U_{\rm c}$ the influence 
of the high-energy Hubbard bands on the low-energy (quasiparticle) 
peak is negligible. This can be specified as follows:
Within the DMFT the Hubbard model is self-consistently mapped 
onto a single-impurity Anderson model (SIAM). In the effective SIAM,
we divide the conduction-electron degrees of freedom 
in a high-energy part $H_{\rm high}$ (the Hubbard bands) and 
a low-energy part $H_{\rm low}$ (the quasiparticle peak). 
The Hamiltonian of the effective SIAM is then written as
\begin{equation}
   H_{\rm SIAM} = H_{\rm high} + H_{\text{high-imp}} + 
           H_{\rm imp} + H_{\rm low} + H_{\text{low-imp}} \: ,
\end{equation}
where the coupling of the impurity $H_{\rm imp} \equiv \sum_{\sigma} 
\epsilon_d d^\dagger_{\sigma} d_{\sigma} + U d^\dagger_{\uparrow} 
d_{\uparrow} d^\dagger_{\downarrow} d_{\downarrow}$ to the high- 
(low-)energy part is denoted as $H_{\text{high-imp}}$ 
($H_{\text{low-imp}}$). The first approximation is then 
to neglect the terms $H_{\rm high}$ and $H_{\text{high-imp}}$. 

To motivate this step, consider the on-site Green function of the
Hubbard model $G(\omega)$. Via the DMFT self-consistency condition,
$G(\omega)$ defines an effective SIAM. For $U\mapsto U_{\rm c}$ one indeed 
finds (e.~g.\ within the NRG) that it makes no significant difference 
for the low-energy part of the solution of the resulting SIAM whether 
the full $G(\omega)$ is considered or the Green function with the Hubbard 
bands removed. This means that in the iterative solution of the DMFT 
equations, the low-energy peak basically reproduces itself, and the 
high-energy degrees of freedom are rather unimportant. 

Alternatively, the first approximation can be characterized as follows: 
Let us (for a moment) look at the insulating solution for $U=U_{\rm c}$. 
Here the low-energy degrees of freedom are absent ($H_{\text{low-imp}}$, 
$H_{\rm low} = 0$) and the approximation reads:
$H_{\rm high} + H_{\text{high-imp}} + H_{\rm imp} \mapsto H_{\rm imp}$.
The impurity spectral function for the left-hand side is given by 
two Hubbard bands centered at $\pm U/2$ while it is given by two 
$\delta$-peaks at $\pm U/2$ for the right-hand side. So we can state 
that in step (i) of the approximation the finite bandwidth of the 
Hubbard peaks is neglected.

(ii) For $U \mapsto U_{\rm c}$, the width of the quasiparticle peak 
vanishes. This fact is used for the second approximation: We assume 
that in this limit it is sufficient to describe the low-energy 
degrees of freedom by a single conduction-band level, i.~e.\ 
$H_{\rm low} \mapsto \sum_\sigma \epsilon_c c^\dagger_\sigma c_\sigma$.
Thereby we disregard any internal structure of the quasiparticle 
peak for $U \mapsto U_{\rm c}$. Equivalently, this means that the 
hybridization function can be represented by a single pole at $\omega=0$:
\begin{equation}
  \Delta(\omega) = \frac{\Delta_N}{\omega} \: .
\label{eq:delta}
\end{equation}
A given hybridization function $\Delta(\omega)$ fixes the effective
impurity problem. Due to 
\begin{equation}
\Delta(\omega) = \sum_{k} \frac{V_k^2}{\omega-(\epsilon_k - \mu)} \: ,
\end{equation}
the one-pole structure of $\Delta(\omega)$ corresponds to an $n_s=2$ 
site single-impurity Anderson model (SIAM):
\begin{eqnarray} 
   H_{\rm 2\mbox{-}site} 
   &=& \sum_{\sigma} \epsilon_d d^\dagger_{\sigma} d_{\sigma} 
                   + U d^\dagger_{\uparrow} d_{\uparrow} 
                       d^\dagger_{\downarrow} d_{\downarrow}
\nonumber \\
     &+& \sum_\sigma \epsilon_c c^\dagger_\sigma c_\sigma
      + \sum_\sigma V (d^\dagger_\sigma c_\sigma + \mbox{h.c.}) 
\label{eq:2site} 
\end{eqnarray}
with the hybridization strength $V = \sqrt{\Delta_N}$. 

Combining (i) and (ii) we obtain: $H_{\rm SIAM} \mapsto H_{\rm imp} + 
H_{\rm low} + H_{\text{low-imp}} \mapsto H_{\rm 2\mbox{-}site}$.
With these two approximations we can run through the DMFT self-consistency 
cycle. In the one-pole ansatz (\ref{eq:delta}) for the hybridization function, 
$\Delta_N$ is the weight of the pole. The index $N$ refers to the $N$-th step 
in the iterative solution. Our goal is to calculate $\Delta_{N+1}$.

We restrict ourselves to the manifest particle-hole symmetric case. The 
chemical potential is set to $\mu=U/2$. The on-site energies in (\ref{eq:2site}) 
are thus given by $\epsilon_d = t_{ii} = 0$ and $\epsilon_c=U/2$. 
For the hybridization strength we have $V = \sqrt{\Delta_N} \mapsto 0$ 
as $U\mapsto U_{\rm c}$.
The two-site impurity model is simple enough to be solved analytically 
\cite{Hewson,Lan98}. For small $V$ there are two peaks in the 
impurity spectral function at $\omega \approx \pm U/2$ as well 
as two peaks near $\omega=0$ which can be considered as corresponding
to the quasiparticle resonance of the infinite ($n_s=\infty$) system.
The weight of this ``resonance'' can be read off from the exact solution
\cite{Hewson}; up to second order in $V/U$ and for the particle-hole 
symmetric case it is given by:
\begin{equation}
  z = 2 \cdot \frac{18 V^2}{U^2} 
  = \frac{36}{U^2} \Delta_N \: .
\label{eq:weight}
\end{equation}

In the self-consistent solution, $z$ is also the quasiparticle 
weight which determines the low-energy behavior of the (local)
self-energy of the lattice problem:
\begin{equation}
  \Sigma(\omega) = U/2 + (1-z^{-1}) \omega + {\cal O}(\omega^2) \: .
\label{eq:sigma}
\end{equation}
For a homogeneous lattice and a local self-energy the on-site Green 
function of the Hubbard model can be written as:
\begin{equation}
  G(\omega) = \int d\varepsilon 
  \frac{\rho(\varepsilon)}{\omega - (\varepsilon - \mu ) - \Sigma(\omega)}
\label{eq:green} \: ,
\end{equation}
where $\rho(\varepsilon)$ is the free ($U=0$) density of states.
Using eq.\ (\ref{eq:sigma}) we obtain:
\begin{equation}
  G(\omega) = z \int d\varepsilon 
  \frac{\rho(\varepsilon)}{\omega - z \varepsilon} + 
G^{\rm (incoh.)}(\omega) \: ,
\end{equation}
where the first part represents the coherent part of the 
Green function ($G^{\rm (coh.)}(\omega)$),
and the second (incoherent) part can be disregarded 
for small excitation energies $\omega \mapsto 0$.

The integration can formally be carried out by means of 
a continued-fraction expansion which for a symmetric 
density of states $\rho(\varepsilon)$ reads:
\begin{eqnarray}
  G^{\rm (coh.)}(\omega) &\equiv& z \int d\varepsilon 
  \frac{\rho(\varepsilon)}{\omega - z \varepsilon} 
  = G^{(U=0)}(z^{-1} \omega) 
\nonumber \\
  &=& 1 / (z^{-1}\omega - b_1^2 / (z^{-1}\omega - b_2^2 / \cdots)) \: .
\end{eqnarray}
The expansion coefficients $b_n$ are related to the moments $M_n$ of the
$U=0$ density of states. The first coefficient $b_1$ is given by:
\begin{equation}
  b_1^2 = M_2 = \int d\varepsilon \: \varepsilon^2 \rho(\varepsilon) \: .
\end{equation}
The second moment $M_2$ is easily calculated by evaluating an 
(anti-)commutator of 
the form $\langle [ \,[[c,H_0]_-,H_0]_-,c^\dagger]_+ \rangle$ which 
yields:
\begin{equation}
  M_2 = \sum_j t_{ij}^2 \: .
\label{eq:moment}
\end{equation}
We thus obtain:
\begin{equation}
  G^{\rm (coh.)}(\omega) = 
  \frac{z}{\omega - z^2 M_2 F(\omega)} \: ,
\label{eq:cf1}
\end{equation}
where we have $F(\omega) = 1/\omega + {\cal O}(\omega^{-2})$
for the remainder. 

Starting from eq.\ (\ref{eq:delta}) in the $N$-th step, the DMFT 
self-consistency equation,
\begin{equation}
  \Delta(\omega) = \omega - (\epsilon_d - \mu)  
  - \Sigma(\omega) - G(\omega)^{-1} \: ,
\label{eq:sc}
\end{equation}
yields a new hybridization function $\Delta(\omega)$ for the $(N+1)$-th step. 
With eqs.\ (\ref{eq:sigma}) and (\ref{eq:cf1}) we get:
\begin{equation}
  \Delta(\omega) = z M_2 F(\omega) 
\label{eq:deltacf}
\end{equation}
for low frequencies $\omega \mapsto 0$. Insisting on the one-pole structure,
\begin{equation}
\Delta(\omega) \stackrel{!}{=} \frac{\Delta_{N+1}}{\omega} \: ,
\end{equation} 
for $U\mapsto U_{\rm c}$, we must have $F(\omega) = 1/\omega$. This 
amounts to replacing the coherent part of the on-site Green function
by the simplest Green function with the same moments up to the 
second one. 

From eqs.\ (\ref{eq:weight}) and (\ref{eq:cf1}) we thus have:
\begin{equation}
  \Delta_{N+1} = \frac{36}{U^2} M_2 \Delta_{N} \: .
\label{eq:ldmft}
\end{equation} 
The coefficient of the $(N+1)$-th iteration step is thereby expressed
in terms of the coefficient of the $N$-th step. This is our main result.
For $U=U_{\rm c}$ the DMFT equations are linearized, they are reduced 
to a simple linear algebraic equation which determines the evolution of 
a single parameter ($\Delta_N$) under subsequent iterations.

The linearized mean-field equation (\ref{eq:ldmft}) 
has only one non-trivial solution with 
$\Delta_{N+1}\!=\!\Delta_N$ which occurs 
for $U\!=\!U_{\rm c}$ with $U_{\rm c}^2\!=\!36M_2$.
Any $U\!<\!U_{\rm c}$ gives $\Delta_{N+1}/\Delta_N>1$, so that  
$\Delta_N$ increases exponentially with iteration number. This indicates 
the breakdown of the one-pole approximation. For any $U>U_{\rm c}$ the weight
$\Delta_N$ decreases exponentially with increasing iteration number.
This corresponds to the vanishing of the quasiparticle peak in the insulating
regime. Consequently, $U_{\rm c}$ has the meaning of the critical interaction 
for the Mott-Hubbard transition and its value is given by:
\begin{equation} 
U_{\rm c} = 6 ~
\sqrt{\int {\rm d} \varepsilon \: \varepsilon^2 \rho(\varepsilon)}  
 = 6 ~ \sqrt{\sum_j t_{ij}^2} \label{eq:Uc} \: .
\label{eq:result}
\end{equation} 
For a lattice with nearest-neighbor coordination number $q$ and 
hopping integral $t=|t_{ij}|$ between nearest-neighbors $i$ and $j$, 
we have: $U_{\rm c} = 6  t ~ \sqrt{q}$.

The result (\ref{eq:result}) has been derived within the DMFT which 
becomes exact in the limit $q \mapsto \infty$.
With the usual scaling for the hopping integral, $t=t^\ast / \sqrt{q}$ 
and $t^\ast = \mbox{const.}$ \cite{MVPRLdinfty}, we have:
\begin{equation} 
  U_{\rm c} = 6 t^\ast \: .
\label{eq:result_inf}
\end{equation}
However, eq.\ (\ref{eq:result}) may also be used for finite-dimensional 
systems where the DMFT is effectively the approximation of a purely local 
self-energy functional.

\section{discussion} 

\subsection{Bethe and hypercubic lattice}

For the Bethe lattice with infinite coordination number and scaling
$t=t^\ast / \sqrt{q}$, the free bandwidth is given by $W\!=\!4t^\ast$.
So we expect from eq.\ (\ref{eq:result_inf}) the Mott-Hubbard
metal-insulator transition to occur at $U_{\rm c}\!=\!1.5W$. This result
is in very good agreement with the result from the
Projective Self-consistent Method (PSCM) \cite{Moeller,Georges}
$U_{\rm c,PSCM}\!\approx\! 1.46W$ and with recent calculations 
using the Numerical
Renormalization Group (NRG) method \cite{NRG_new} which yield
$U_{\rm c,NRG}\!\approx\! 1.47W$. It also agrees well with the value of
     $U_{\rm c}\!\approx 1.51W$ obtained in the NRG calculations of 
     Shimizu and Sakai \cite{Shi95}. The earlier IPT result 
$U_{\rm c,IPT}\!\approx\! 1.65W$ \cite{Georges} overestimates the critical $U$
as compared to the other, non-perturbative methods. 
The Random Dispersion Approximation
(RDA) \cite{NG} predicts a considerably lower critical value
$U_{\rm c,RDA}\approx W$. The origin of this discrepancy, however, is
presently not clear.

On the infinite-dimensional hypercubic lattice with the scaling 
$t=t^\ast / \sqrt{q}$, we expect the metal-insulator transition
to occur at $U_{\rm c}=6t^\ast$. Again, this agrees well with the
NRG calculations \cite{NRG_new} where the value 
$U_{\rm c}\approx 5.80t^\ast$ has been found.

The existence of a metal-insulator transition in the hypercubic 
lattice at a finite $U_{\rm c}$ is not at all clear, considering 
the fact that the free density of states is Gaussian, i.e.\ has no 
cutoff. In any case, the actual bandwidth (which is infinite for a 
Gaussian density of states) cannot play a role for the value of 
$U_{\rm c}$. It is much more plausible that it is the effective
bandwidth (which is proportional to $\sqrt{\int {\rm d} \varepsilon 
\rho(\varepsilon) \varepsilon^2}$) that has to be taken as a measure 
for $U_{\rm c}$. 

Our analysis also shows that $U_{\rm c}$ is roughly independent 
of the details of the lattice structure and only depends on the 
{\em local} quantity $\sum_j t_{ij}^2$.
This result can quite naturally be understood when the electrons are
considered as getting localized at the transition. In this case 
the electrons would only see their immediate surrounding which is 
the same for both the infinite
dimensional Bethe and the hypercubic lattice.

\subsection{Two-sublattice model}

Let us now work out the predictions of the linearized DMFT for 
{\em inhomogeneous} lattices, i.~e.\ lattices with reduced (translational) 
symmetries.
The presumably simplest but non-trivial case is a Bethe lattice that consists
of two non-equivalent sublattices $Q_1$ and $Q_2$ where each site of $Q_\alpha$
has $q_\alpha$ nearest neighbors that belong to $Q_{\overline{\alpha}}$
(with $\overline{\alpha} = 2$ for $\alpha=1$ and $\overline{\alpha} = 1$ 
for $\alpha=2$). We consider the limit of infinite coordination numbers
$q_1,q_2 \mapsto \infty$ with $0 < q_1/q_2 < \infty$. As for the 
homogeneous case $q_1=q_2$, it can be shown that the Hubbard model on the 
inhomogeneous lattice remains well-defined and non-trivial if the hopping
integral is scaled appropriately, e.~g.\ $t=t^\ast / \sqrt{q_1+q_2} = 
t^{\ast \ast} / \sqrt{q_1}$ with $t^\ast, t^{\ast \ast} = \mbox{const.}$
As a consequence, the self-energy $\Sigma_\alpha(\omega)$ is local but
sublattice dependent. The lattice problem can be mapped onto two impurity
models that are characterized by hybridization functions $\Delta_\alpha(\omega)$.
The DMFT self-consistency equations read:
\begin{equation}
  \Delta_\alpha(\omega) = \omega - (\epsilon_d - \mu) - \Sigma_\alpha(\omega) 
  - G_\alpha(\omega)^{-1} \: ,
\label{eq:scih}
\end{equation}
where $G_\alpha(\omega)$ is the on-site Green function for a site $i$ within the
sublattice $\alpha$. One easily verifies that the free ($U=0$) local density of
states on each sublattice is symmetric and that $\mu=U/2$ at half-filling.
Furthermore, with $q_1,q_2\mapsto \infty$, we obtain from the lattice Dyson 
equation:
\begin{equation}
  G_\alpha(\omega)^{-1} = \omega + \mu - \Sigma_\alpha(\omega) 
  - q_\alpha t^2 G_{\overline{\alpha}}(\omega) \: .
\label{eq:bethedyson}
\end{equation}

The linearized DMFT for $U\mapsto U_{\rm c}$ iterates the one-pole ansatz 
$\Delta_\alpha(\omega) = \Delta_N^{(\alpha)} / \omega$. From eqs.\ 
(\ref{eq:scih}) and (\ref{eq:bethedyson}) we have
$\Delta_\alpha(\omega) = q_\alpha t^2 G_{\overline{\alpha}}(\omega)$
for $\alpha = 1,2$. This implies that the quasiparticle peak for the 
sublattice $\overline{\alpha}$ with weight 
$z_{\overline{\alpha}}=(36/U^2)\Delta_N^{(\overline{\alpha})}$ generates
a corresponding peak in $\Delta_\alpha(\omega)$ with the weight
$\Delta_{N+1}^{(\alpha)} = q_\alpha t^2 z_{\overline{\alpha}}$.
We thus get:
\begin{equation}
  \Delta^{(\alpha)}_{N+1} = \sum_\beta K_{\alpha \beta}(U) \:
  \Delta^{(\beta)}_{N} \: ,
\label{eq:lindmft}
\end{equation}
where the $2\times 2$ matrix ${\bf K}(U)$ is defined as:
\begin{equation}
  {\bf K}(U) = \frac{36 \, t^2}{U^2} \left( 
  \begin{array}{cc}
  0   & q_1 \\
  q_2 & 0   \\
  \end{array} \right) \: .
\label{eq:k1}
\end{equation}
A fixed point of ${\bf K}(U)$ corresponds to a self-consistent solution. 
Let $\lambda_r(U)$ denote the eigenvalues of ${\bf K}(U)$. We can distinguish 
between two cases: If $|\lambda_r(U)| < 1$ for $r=1$ and $r=2$, there is the
trivial solution $\lim_{N\mapsto \infty} \Delta^{(\alpha)}_{N} = 0$ only 
(insulating solution for $U>U_{c}$). On the other hand, if there is at least 
one $\lambda_r(U)>1$, $\Delta^{(\alpha)}_{N}$ diverges exponentially as 
$N \mapsto \infty$ (metallic solution for $U<U_{c}$). The critical interaction 
is thus determined via the maximum eigenvalue by the condition:
\begin{equation}
  \lambda_{\rm max}(U_{c}) = 1 \: .
\label{eq:cond}
\end{equation}
This yields:
\begin{equation}
  U_{c} = 6t~ \sqrt[4]{q_1q_2}  \: ,
\end{equation}
i.~e.\ the geometrical mean of the critical interactions of two homogeneous
Bethe lattices with coordination numbers $q_1$ and $q_2$, respectively. The 
result recovers the homogeneous case $q_1=q_2$ and correctly gives $U_{\rm c}=0$ 
for the atomic limit $q_1\mapsto 0$ or $q_2\mapsto 0$.

The analysis can be generalized straightforwardly to an arbitrary number of $s$
sublattices $Q_1 , \dots , Q_s$. We consider a Bethe lattice where each site of
the sublattice $Q_\alpha$ has $(q_\alpha - 1)$ nearest neighbors belonging to
the sublattice $Q_{\alpha_+}$ and one nearest neighbor in the sublattice 
$Q_{\alpha_-}$ where $\alpha_{\pm} \equiv \alpha \pm 1$ except for $\alpha = s$
(here $\alpha_+ \equiv 1$) and $\alpha = 1$ ($\alpha_- \equiv s$). In the limit
$q_\alpha \mapsto \infty$ with fixed pairwise ratios $0<q_\alpha/q_\beta<\infty$,
we have:
\begin{equation}
  G_\alpha(\omega)^{-1} = \omega + \mu - \Sigma_\alpha(\omega) 
  - q_\alpha t^2 G_{\alpha_+}(\omega) \: ,
\end{equation}
and the argument proceeds as above. We finally arrive at the mean-field
equation (\ref{eq:lindmft}) with ${\bf K}(U)$ being an $s$-dimensional matrix
with $s$ non-zero elements:
\begin{equation}
  {\bf K}(U) = \frac{36 \, t^2}{U^2} \left( 
  \begin{array}{ccccc}
  0  & q_1 &      &    &         \\
     & 0   & q_2  &    &         \\
     &     & 0    & .. &         \\
     &     &      & .. & q_{s-1} \\
 q_s &     &      &    & 0       \\
  \end{array} \right) \: .
\label{eq:k2}
\end{equation}
This implies:
\begin{equation}
  U_{c} = 6t~\left( \prod_{\alpha=1}^s {q_\alpha} \right)^{1/2s} \: .
\label{eq:ucs}
\end{equation}
Again, this is plausible since $q_\alpha=0$ for any $\alpha$ would mean to cut
the lattice into unconnected pieces of finite size, and the Mott transition 
becomes impossible ($U_{\rm c}=0$).

Also the $s\mapsto \infty$ limit of eq.\ (\ref{eq:ucs}) is meaningful: 
Consider e.~g.\ $q_{\alpha=1} \ne q \equiv q_2 = q_3 = \cdots$. This 
describes a Bethe lattice with coordination number $q$ for all sites except 
for one distinguished impurity site with coordination number $q_1$. 
As expected physically, 
$U_{\rm c}$ is unaffected by the presence of the impurity. 
Furthermore, in any case where one changes the number of nearest neighbors 
of a {\em finite} number of sites only, 
the value for $U_{\rm c}$ remains unchanged.

\subsection{General inhomogeneous Bethe lattice}

We finally tackle the ``inverse'' problem: Given a matrix $\bf K$, is there a
realization of a (Bethe) lattice such that the critical interaction is
determined by the maximum eigenvalue of $\bf K$? For this purpose we 
consider the Hubbard model with nearest-neighbor hopping on a
{\em general} inhomogeneous Bethe lattice where each site $i$ may have a 
different coordination number. Remaining spatial symmetries are accounted
for by classifying the lattice sites into sublattices $Q_\alpha$ that 
consist of equivalent sites only.
By $q_{\alpha \beta}$ we denote the number of nearest neighbors of a site
$i \in Q_\alpha$ that belong to the sublattice $Q_\beta$. 
We are interested in the limit $q_{\alpha \beta} \mapsto \infty$ with
$0 < q_{\alpha \beta} / q_{\gamma \delta} < \infty$ since this implies
a local but $\alpha$-dependent self-energy $\Sigma_\alpha(\omega)$.
Within the DMFT this Hubbard model is mapped onto impurity models which are
labeled by the sublattice index $\alpha$. The self-consistency conditions
are given by eq.\ (\ref{eq:scih}).

Let $G_{\alpha}^{(0)}(\omega) \equiv G_\alpha(\omega)$ 
be the on-site Green function for a site $i$ in $Q_\alpha$,
and $G^{(n)}_{\alpha_n \alpha_{n-1} \cdots \alpha_1 \alpha}(\omega) 
= \langle \langle c_{i\sigma} ; c_{j\sigma}^\dagger \rangle \rangle_\omega$
the off-site Green function for $n$-th nearest-neighbor sites 
$i \in Q_{\alpha_n}$ and $j \in Q_{\alpha}$. $G^{(n)}$ depends 
on the sublattice indices that are met along the (unique)
path from $j$ to $i$.
Via its equation of motion, $G_{\alpha}^{(0)}$ 
couples to the nearest-neighbor off-site
Green function $G^{(1)}_{\alpha_1\alpha}$:
\begin{equation}
  (\omega + \mu - \Sigma_\alpha(\omega)) G_{\alpha}^{(0)}(\omega) 
  = 1 + t \sum_{\alpha_1} q_{\alpha\alpha_1} 
  G^{(1)}_{\alpha_1\alpha}(\omega) \: .
\label{eq:eom1}
\end{equation}
For a Bethe lattice the nearest-neighbor Green function
$G^{(1)}_{\alpha_1\alpha}$ can only couple to $G^{(2)}_{\alpha_2\alpha_1\alpha}$ 
and to $G_{\alpha}^{(0)}$ again. More generally, the equation of motion for the 
$n$-th nearest-neighbor off-site Green function reads:
\begin{eqnarray}
  (\omega + \mu - \Sigma_{\alpha_n}(\omega)) 
  G^{(n)}_{\alpha_n \cdots \alpha_1 \alpha}(\omega) &=&
  t G^{(n-1)}_{\alpha_{n-1} \cdots \alpha_1 \alpha}(\omega) 
\nonumber \\ \mbox{} && \hspace{-20mm} + 
  t \sum_{\alpha_{n+1}} q_{\alpha_n \alpha_{n+1}} 
  G^{(n+1)}_{\alpha_{n+1} \cdots \alpha_1 \alpha}(\omega) \: ,
\label{eq:eom2}
\end{eqnarray}
where in the second term on the r.h.s.\ we have used the approximation 
$q_{\alpha \beta} - 1 \approx q_{\alpha\beta}$ which becomes exact
in the limit of infinite coordination numbers. The infinite series defined
by eqs.\ (\ref{eq:eom1}) and (\ref{eq:eom2}) can formally be summed up.
This yields:
\begin{equation}
  G_\alpha(\omega)^{-1} = \omega + \mu - \Sigma_\alpha(\omega) 
  - t^2 \sum_\beta q_{\alpha\beta} G_{\beta}(\omega) \: .
\end{equation}

The mean-field equation of the linearized DMFT thus has again the form
(\ref{eq:lindmft}) where the $K$-matrix is given by:
\begin{equation}
  K_{\alpha\beta}(U) = \frac{36t^2}{U^2} q_{\alpha \beta} \: .
\label{eq:kcoord}
\end{equation}
The critical interaction is given by $6t$ times the maximum eigenvalue 
of the coordination-number matrix $\bf q$. Note that for a general
(non-symmetric), irreducible
matrix with non-negative elements, the eigenvalue with 
maximum absolute value is real and non-negative (Perron-Frobenius
theorem \cite{pf}). 
We thus conclude that any quadratic matrix $\bf K$ with non-negative 
elements can be related to the Mott transition on a
Bethe lattice with certain (infinite) coordination numbers.

\subsection{Hypercubic films}

As a more realistic example for the Mott transition on an 
inhomogeneous lattice we consider a hypercubic Hubbard film.
A $D$-dimensional film is built up from a number $d$ 
of $(D-1)$-dimensional ``layers''. For a hypercubic film 
these layers are cut out of the usual $D$-dimensional hypercubic
lattice. A set of Miller indices $[x_1,x_2, \dots ,x_D]$
characterizes the film-surface normal direction. The most simple 
films are those with low-index surfaces given by 
$x_1 = \cdots = x_r = 1$ and $x_{r+1} = \cdots = x_D = 0$. 
For any site in the film except for sites at the film surfaces
there are $q_{1}=2D-2r$ nearest neighbors within the same 
layer and $q_{2}=r$ nearest neighbors in each of the 
adjacent layers; the total coordination number is 
$q=q_{1}+2q_{2}=2D$.

For $D\mapsto \infty$ the Hubbard model is well defined with the usual 
scaling of the hopping $t = t^\ast / \sqrt{2D}$, the self-energy becomes 
local but layer dependent and dynamical mean-field theory is exact 
\cite{PN99}. The lattice problem is mapped onto a set of $d$ impurity
problems. The DMFT self-consistency conditions are given by eq.\ 
(\ref{eq:scih}) where the index $\alpha$ now has to be interpreted
as the layer index: $\alpha=1,\dots,d$.

The linearized DMFT can be developed as in Sec.\ II. Equation 
(\ref{eq:green}), however, is no longer valid and must be replaced 
by the Dyson equation
corresponding to the given film geometry. The coherent part of the on-site
Green function for a site in the layer $\alpha$ is given by:
\begin{equation}
  G_\alpha(\omega) = z_\alpha \widetilde{G}_\alpha(\omega)
  = \frac{z_\alpha}{\omega - \widetilde{M}_2^{(\alpha)} F_\alpha(\omega)}
\end{equation}
where $z_\alpha$ is the layer-dependent quasiparticle weight,
$z_\alpha = (1-\partial \Sigma_\alpha(i0^+)/\partial \omega)^{-1}$ and 
$\widetilde{G}_\alpha(\omega)$ is the on-site element of the free ($U=0$)
Green function but calculated for the renormalized hopping
$t_{ij} \mapsto \sqrt{z_i}~t_{ij} \sqrt{z_j}$ with $z_i=z_\alpha$ for 
a site $i$ in the layer $\alpha$. In the expression on the right, 
$\widetilde{M}_2^{(\alpha)}$ denotes the corresponding second moment 
which is calculated as $\widetilde{M}_2^{(\alpha)}=z_\alpha(q_1 z_\alpha+
q_2 z_{\alpha-1} + q_2 z_{\alpha+1})t^2$. This means that the 
linearized mean-field equation has again the form (\ref{eq:lindmft}) with
the following $d$-dimensional tridiagonal matrix:
\begin{equation}
  {\bf K}(U) = \frac{36 \, t^2}{U^2} \left( 
  \begin{array}{cccc}
  q_{1} & q_{2} &       &         \\
  q_{2} & q_{1} & q_{2} &         \\
        & q_{2} & q_{1} & ..      \\
        &       & ..    & ..      \\
  \end{array} \right) \: .
\label{eq:k4}
\end{equation}
Its eigenvalues are the zeros of the $d$-th degree Chebyshev polynomial
of the second kind \cite{steeb}. From the maximum eigenvalue we obtain:
\begin{equation}
  U_{c} = 6 t \: \sqrt{q_1+2q_2 \cos\left(\frac{\pi}{d+1}\right)} \: .
\label{eq:ucfilm}
\end{equation}
Equation (\ref{eq:ucfilm}) describes the complete thickness and geometry
dependence of the critical interaction for the Mott-Hubbard transition in
hypercubic Hubbard films. 

In the limit of thick films $d\mapsto\infty$ one
recovers the bulk value $U_{\rm c}= 6t \sqrt{q_1+2q_2}$. For $d<\infty$ the 
critical interaction not only depends on the film thickness $d$ but also
on the geometry of the film surface which is characterized by $r$.
Varying $r$ we can pass continuously from the most closed ($r=1$) to the 
most open ($r=D$) surface geometry. For $r=1$, i.~e.\ a (1000...) film 
surface, a site in the topmost layer has $q_{\rm S}=q_1+q_2=2D-1$ nearest 
neighbors to be compared with $q=2D$ in the bulk. For $D \mapsto \infty$ 
the local environment of the surface sites is essentially the same as in 
the bulk, i.~e.\ surface effects become meaningless. Consequently, we get 
$U_{\rm c} = 6t^\ast$, i.~e.\ the bulk value irrespective of $d$. For $r=D$ 
one obtains the open 
(1111...) film surface. The surface coordination number is reduced 
to $q_{\rm S}=q_2=D$. The critical interaction is 
$U_{\rm c}=6t^\ast \sqrt{\cos (\pi / (d+1))}$ which is smaller than $6t^\ast$
for any $d$.

Equation (\ref{eq:ucfilm}) can also be applied to finite-dimensional films
($D<\infty$) if one additionally assumes the local approximation for the
self-energy functional to hold. For $D=3$ simple-cubic films with a
thickness ranging from $d=1$ up to $d=8$ and for sc(100), sc(110) and sc(111)
film surfaces, the prediction (\ref{eq:ucfilm}) of the linearized DMFT
has been tested in Ref.\ \cite{PN99} by comparing with the results for 
$U_{\rm c}$ of a fully numerical evaluation of the DMFT equations using 
the exact diagonalization of small impurity models ($n_s=8$).
It is found that the linearized DMFT qualitatively and -- as far as can be
judged from the numerical evaluation -- also quantitatively predicts the 
correct geometry and thickness dependence of $U_{\rm c}$ \cite{PN99}.

\subsection{Critical exponent and critical profiles} 

So far the discussion was restricted to the calculation of the
critical value $U_{\rm c}$ which is derived from a {\em linear}
homogeneous mean-field equation [eqs.\ (\ref{eq:ldmft}) and 
(\ref{eq:lindmft})]. To determine the critical behaviour of 
the quasiparticle weight $z$ for $U\mapsto U_{\rm c}$, one has to 
go beyond the linearized DMFT. For this purpose a simple 
generalization of the arguments in Sec.\ II is necessary.

We replace the second-order result for the quasiparticle weight 
$z\!=\!36 V^2/U^2$ [eq.\ (\ref{eq:weight})] by the result up to 
fourth order in $V/U$:
\begin{equation}
   z = 36 \frac{V^2}{U^2} \left(
              1 - 44 \frac{V^2}{U^2} \right) \: .
\end{equation}
With the same steps as before, one arrives at:
\begin{equation}
  \Delta_{N+1} = \left(
              1 - 44 \frac{\Delta_{N}}{U^2} \right)
\frac{36}{U^2} M_2 \Delta_{N} \: ,
\label{eq:Del_new}
\end{equation}
which is a non-linear equation for the ``mean field'' $\Delta$.
The self-consistency requires 
$\Delta_{N+1}\!=\!\Delta_{N}\!=\!\Delta$. Solving for $\Delta$
yields:
\begin{equation}
   \Delta = \frac{1}{22} U_{\rm c} \left( U_{\rm c} - U
            \right) \: ,
\end{equation}
where we have already expanded the right hand side in
powers of $(U_{\rm c} - U)$. The result for the quasiparticle
weight near $U_{\rm c}$ is
\begin{equation}
    z = \frac{18}{11} \frac{U_{\rm c} - U}{U_{\rm c}} \: .
\label{eq:zcrit}
\end{equation}
This equation is, of course, only valid for $U<U_{\rm c}$. 
We obtain a {\it linear} vanishing of the quasiparticle 
weight near the metal-insulator transition as in the
Brinkman-Rice approach \cite{BR70} and the PSCM \cite{Moeller}.

For the Hubbard model on an inhomogeneous lattice, the self-energy
and thus the quasiparticle weight is site or sublattice dependent:
$z_\alpha=(1-\partial \Sigma_\alpha(i0^+) / \partial \omega )^{-1}$.
Within the linearized DMFT [eq.\ (\ref{eq:lindmft})], the critical 
interaction is determined from the largest eigenvalue 
$\lambda_{\rm max}(U_{c}) = 1$ in the eigenvalue problem
\begin{equation}
  z_\alpha = \sum_\beta K_{\alpha \beta}(U_{\rm c}) ~ z_\beta \: .
\end{equation}
The corresponding eigenvector $z_\alpha=z_\alpha(U_{\rm c})$ describes 
the critical {\em profile} of the quasiparticle weight. The profile 
is uniquely determined up to a normalization constant. While 
$z_\alpha(U) \mapsto 0$ for each $\alpha$ as $U \mapsto U_{\rm c}$, 
the ratios $z_\alpha(U_{\rm c})/z_\beta(U_{\rm c})$ remain to be non-trivial.
For example, in the two-sublattice model characterized by eq.\ 
(\ref{eq:k1}), the critical profile is given by 
$z_1(U_{\rm c})/z_2(U_{\rm c}) = \sqrt{q_1/q_2}$.

To determine the critical behaviour of the $\alpha$-dependent
quasiparticle weight for $U\mapsto U_{\rm c}$ but $U<U_{\rm c}$, we again 
have to expand up to fourth order in $V_\alpha/U$. This yields
the following mean-field equation:
\begin{equation}
  z_\alpha(U) = \sum_\beta K_{\alpha \beta}(U) ~ z_\beta(U) 
           - \frac{11}{9} z_\alpha^2(U)
  \: .
\end{equation}
For the two-sublattice model we obtain:
\begin{equation}
  z_{1,2}(U) = \frac{36}{11} \frac{\sqrt{q_{1,2}}}{\sqrt{q_1}+\sqrt{q_2}} ~
  \frac{U_{\rm c} - U}{U_{\rm c}} \: .
\end{equation}
For $q_1=q_2$ this result reduces to eq.\ (\ref{eq:zcrit}).

\section{summary} 

We have discussed a ``linearized'' version of the Dynamical
Mean Field Theory which allows for the analytical calculation of
the critical interaction for the Mott-Hubbard metal-insulator
transition at $T\!=\!0$. The main result is:
\begin{equation} 
U_{\rm c} = 6 ~
\sqrt{\int {\rm d} \varepsilon \: \varepsilon^2 \rho(\varepsilon)}  
 = 6 ~ \sqrt{\sum_j t_{ij}^2} \: ,
\end{equation} 
which shows that it is the second moment of the non-interacting
density of states which determines $U_{\rm c}$.
This is contrary to the Hubbard-III approximation where the 
critical interaction is determined by the free bandwidth $W$ 
and to the Brinkman-Rice approach where $U_{\rm c}$ is given in terms 
of the $U=0$ kinetic energy $e_0$.
The values for $U_{\rm c}$ obtained with the linearized DMFT 
have been compared with the available results from numerical
solutions of the full DMFT equations, and a good agreement is 
found.

The linearized DMFT is of course not able to answer detailed
questions about the nature of the Mott-Hubbard transition,
such as the existence or absence of a hysteresis, the order
of the transition, etc. Its advantage is that it can be easily
generalized to a variety of geometries. To obtain the critical
interaction, it is sufficient to find the maximum eigenvalue of 
the respective coordination-number matrix, the dimension of which
is determined by the remaining spatial symmetries 
(see e.g.\ eq.\ (\ref{eq:kcoord})). The analytical results
for e.~g.\ the metal-insulator transition in thin Hubbard films
have been checked against numerical solutions of the full DMFT 
equations, and the geometry dependence has been found to be 
essentially the same, in the DMFT and in the linearized DMFT.

The main approximation which our approach is based on, is the 
one-pole structure of the effective hybridization function. It 
is obvious that the linearized DMFT can in principle be extended 
by taking into account more states for the effective conduction 
band (in the quasiparticle peak and/or the Hubbard bands). However, 
its main advantage -- the possibility to obtain analytical results 
for $U_{\rm c}$ -- would then be immediately lost. 

It would be very interesting to see whether experiments on 
Mott-Hubbard systems in different geometries will show similar 
trends as predicted by the linearized DMFT.

\section*{acknowlegdements}

We would like to thank A.~C.\ Hewson, W.~Nolting, and Th.~Pruschke for 
discussions. R.~B. thanks the Max-Planck-Institut f\"ur Physik komplexer 
Systeme in Dresden for hospitality while part of this work was done. 
The support by the Deutsche Forschungsgemeinschaft within the SFB~290
is gratefully acknowledged.

\end{document}